\documentclass[floatfix,twocolumn,showpacs,prl,amsmath]{revtex4}
\usepackage{graphicx}
\usepackage{bm}
\usepackage{amssymb}
\usepackage{dcolumn}
\usepackage{subfigure}
\usepackage{threeparttable}
\usepackage{multirow}
\usepackage[usenames,dvipsnames]{color}
\usepackage{mathtools}

\begin{document}
\newcommand{\vn}[1]{{\boldsymbol{#1}}}
\newcommand{\vht}[1]{{\boldsymbol{#1}}}
\newcommand{\matn}[1]{{\bf{#1}}}
\newcommand{\matnht}[1]{{\boldsymbol{#1}}}
\newcommand{\bege}{\begin{equation}}
\newcommand{\ee}{\end{equation}}
\newcommand{\bal}{\begin{aligned}}
\newcommand{\defbar}{\overline}
\newcommand{\SM}{\scriptstyle}
\newcommand{\eal}{\end{aligned}}
\newcommand{\torkance}{t}
\newcommand{\udot}{\overset{.}{u}}
\newcommand{\exponential}[1]{{\exp(#1)}}
\newcommand{\phandot}[1]{\overset{\phantom{.}}{#1}}
\newcommand{\phandag}{\phantom{\dagger}}
\newcommand{\Trace}{\text{Tr}}
\newcommand{\Bxc}{\Omega}
\newcommand{\mubo}{\mu_{\rm B}^{\phantom{B}}}
\newcommand{\rmd}{{\rm d}}
\newcommand{\rme}{{\rm e}}
\setcounter{secnumdepth}{2}
\title{Charge pumping driven by the laser-induced dynamics of the exchange splitting}
\author{Frank Freimuth}
\email[Corresp.~author:~]{f.freimuth@fz-juelich.de}
\author{Stefan Bl\"ugel}
\author{Yuriy Mokrousov}
\affiliation{Peter Gr\"unberg Institut and Institute for Advanced Simulation,
Forschungszentrum J\"ulich and JARA, 52425 J\"ulich, Germany}
\date{\today}
\begin{abstract}
We show that electric currents are induced in inversion asymmetric
ferromagnets if the exchange splitting varies in time after excitation
by laser pulses. We interpret this phenomenon as the
magnetic variant of the inverse Edelstein effect.
Based on \textit{ab initio} calculations we determine
the size of this effect in Co/Pt(111) and Mn/W(001) magnetic bilayer
systems and investigate its dependence on the magnetization direction.
The comparison of our theoretical results to experiments
measuring the THz signal emitted after laser excitation suggests that
ultrafast demagnetization in 3$d$ transition metal ferromagnets 
is dominated by transverse spin fluctuations
rather than by a reduction of the exchange splitting.
\end{abstract}


\maketitle
\section{Introduction}
Optical excitation of magnetic heterostructures
such as Fe/Au or Co/Pt by femtosecond (fs) laser
pulses triggers in-plane subpicosecond 
electric currents~\cite{thz_spin_current_kampfrath,Huisman_2016,seifert_THz_emitter}.
Two distinct mechanisms of this effect have been identified already:
First, a superdiffusive spin current is launched, which propagates from
the magnetic layer into the nonmagnetic 
metal~\cite{Malinowski_ultrafast_demag_direct_spin_transfer,
battiato_superdiffusive_spin_transport_ultrafast_demagnetization,
melnikov_AuFeMgO_PhysRevLett.107.076601}.
The generation of this superdiffusive spin current can be interpreted as an
ultrafast nonequilibrium variant of the spin-dependent Seebeck effect triggered by the
laser-induced heating~\cite{PhysRevB.86.024404}.
The inverse spin Hall effect (ISHE) converts this subpicosecond spin current 
into a charge
current pulse flowing parallel to the bilayer 
interface~\cite{thz_spin_current_kampfrath,seifert_THz_emitter}.
Second, when a circularly polarized laser beam is used,
an additional effect can be observed:
The magnetization of the magnetic layer tilts
because of the effective magnetic field generated by the 
inverse Faraday effect or by the optical spin-transfer 
torque~\cite{Kimel_ultrafast_control_magnetization,nemec_ostt,lasintor}. 
Via charge pumping due to the inverse spin-orbit 
torque (ISOT)~\cite{PhysRevB.92.064415,PhysRevB.91.214401,charge_pumping_Ciccarelli} 
the tilting magnetization induces an interfacial electric
current pulse~\cite{Huisman_2016}.

These in-plane subpicosecond photocurrents can be quantified
contactlessly by measuring the resulting emission of terahertz (THz) electromagnetic
radiation~\cite{thz_spin_current_kampfrath,Huisman_2016}.
By optimizing the composition and layer thicknesses of the
magnetic heterostructures one can even achieve THz radiation that
is sufficiently strong to make
photocurrents in magnetic heterostructures 
attractive for application in table-top ultrabroadband THz emitters~\cite{seifert_THz_emitter}.
Additionally,
the photocurrents might provide a new and complementary tool 
to investigate magnetization dynamics at subpicosecond time scales, where other
magneto-optical probing techniques 
face
difficulties~\cite{PhysRevB.61.14716,ultrafast_magneto_optics_nickel,
simultaneous_thz_moke_huisman}.

Another effect triggered by fs laser pulses is ultrafast 
demagnetization~\cite{PhysRevLett.76.4250}.
Several experiments and theories suggest that ultrafast demagnetization 
is accompanied by a break-down of the local magnetic moments and 
by a collapse of the exchange 
splitting~\cite{PhysRevLett.111.167204,exchange_interaction_quenching,PhysRevB.91.014425,demagnetization_tddft}.
Since the electric conductivity tensor is affected by such
electronic structure changes one can use measurements of the electric conductivity tensor in order to
assess to which extent ultrafast demagnetization is accompanied by a reduction of the 
exchange 
splitting~\cite{collapsed_vs_collective}. In 
such experiments only small changes of the conductivity
have been found in 50nm thick Fe(001)/MgO(001) films. 
Furthermore, for unsupported Ni monolayers,  
it is found from exact time propagation of a many-electron small-cluster model  
that laser-induced demagnetization arises dominantly from fluctuations
of the orientations of the local
magnetic moments, while the local spin polarization is not 
reduced significantly~\cite{ufd_pastor}. 
In contrast, in the strong-field limit,
which can be realized by reducing the laser-pulse duration while
keeping the fluence constant, it has been predicted that the
local magnetic moments are reduced and that no fluctuations of the
orientations of the local
magnetic moments are excited~\cite{demagnetization_tddft}. 
Additionally, it is expected that the extent to which a reduction 
of the exchange splitting contributes to
ultrafast demagnetization is material-dependent:
When the local magnetic moments are particularly stable, a collapse
of the exchange splitting is less likely than fluctuations of the
orientations of the local
magnetic moments~\cite{ufd_pastor}. 
 
Moreover, whether ultrafast demagnetization is dominated by
transverse fluctuations or by a reduction of the local magnetic moments
might also depend on the geometry. Superdiffusive spin currents, 
which carry magnetization away from the magnetic layers,
provide a nonlocal transport contribution to ultrafast demagnetization in metallic
heterostructures, which adds to the demagnetization due to
the local spin-flip processes mediated by 
the spin-orbit interaction (SOI)~\cite{Malinowski_ultrafast_demag_direct_spin_transfer,
battiato_superdiffusive_spin_transport_ultrafast_demagnetization,
PhysRevB.86.024404,PhysRevB.92.174410,demagnetization_tddft}.
In heterostructures composed of two Co/Pt multilayers separated by a Ru spacer
the contribution of
the superdiffusive spin currents to the total amount of
demagnetization has been estimated
at 25\%~\cite{Malinowski_ultrafast_demag_direct_spin_transfer}.
In Ni/Ru/Fe heterostructures
30\% of demagnetization of Ni is attributed to
superdiffusive spin currents flowing from
Fe to Ni~\cite{PhysRevB.90.104429}.
The semiclassical theory seems to suggest that demagnetization by
superdiffusive spin currents goes hand in hand with a reduction of
local spin polarization, at least initially. However, it has been
pointed out that even in this case the system will quickly reach
a state where the local magnetic moments are tilted against
each other while they have the same modulus as in 
equilibrium~\cite{PhysRevB.92.064403}.

Since small-angle precession of the 
magnetization at GHz frequencies, i.e., the coherent dynamics of the
\textit{transverse}
degree of freedom, generates measurable electrical
currents~\cite{charge_pumping_Ciccarelli}
as well as heat currents~\cite{itsot,PhysRevB.91.014417} 
in metallic ferromagnets with inversion 
asymmetry, the question 
arises whether also the subpicosecond reduction of the local 
spin polarization, i.e., the dynamics of the
\textit{longitudinal} degree of freedom, induces such
currents. 
The idea that the dynamics of the exchange splitting can 
induce measurable charge or heat currents 
seems realistic in view of the strong formal similarities with the inverse Edelstein effect,
which consists in the conversion of spin currents flowing towards
interfaces or surfaces into
transverse charge currents~\cite{spin_to_charge_rojas_sanchez,inverse_rashba_cubi}. 
In order to calculate the inverse Edelstein effect,
which
results from inversion asymmetry and SOI
at interfaces or surfaces, one can describe the
spin-injection into the interface or surface by a magnetic field that varies 
linearly in time~\cite{PhysRevLett.112.096601}. Experiments 
on the inverse Edelstein effect
are usually performed on nonmagnetic interfaces 
such as Bi/Ag~\cite{spin_to_charge_rojas_sanchez}
and Cu/Bi~\cite{inverse_rashba_cubi}. Here, we are interested
in magnetic bilayer systems such as Co/Pt or Mn/W, where the exchange-splitting is
time-dependent after excitation by a fs laser pulse. Since a time-dependent
magnetic field induces charge currents in Bi/Ag and Cu/Bi, it is natural
to expect that the dynamics of the exchange splitting induces charge currents in
Co/Pt and Mn/W magnetic bilayers as well.
The generation of electric currents by 
time-dependent magnetic fields in nonmagnetic materials is
also known as gyrotropic magnetic effect~\cite{gyrotropic_magnetic_effect}.

If a dynamical exchange splitting drives charge currents, this effect
can be employed in experiments in order to investigate the question 
whether ultrafast demagnetization is dominated by transverse
spin fluctuations or a reduction of the local magnetic moments depending
on the magnetic material, the laser pulse characteristics and the geometry.
In this paper we determine
the charge current driven by a dynamical exchange
splitting in Co/Pt(111) and Mn/W(001) magnetic bilayers based on \textit{ab initio}
density functional theory calculations. This paper is structured as follows:
In section~\ref{sec_formalism} we introduce our computational approach,
which uses the Kubo linear-response formalism in order to calculate the
charge pumping by a dynamical exchange splitting.
In section~\ref{sec_results} we present the results for Co/Pt(111) and Mn/W(001) magnetic bilayers,
we compare to recent experiments measuring the THz emission from similar bilayer systems under
laser irradiation, and we discuss implications of this theory-experiment comparison for the
ultrafast demagnetization process in these bilayer systems.
We conclude with a summary in section~\ref{sec_summary}.
\section{Formalism}
\label{sec_formalism}
Within the local spin density approximation (LSDA)
interacting many-electron systems in equilibrium are described by
the effective single-particle Hamiltonian
\bege\label{eq_ks_hamiltonian}
H(\vn{r})=H_{0}(\vn{r})-\vn{m}\cdot\hat{\vn{M}}\Bxc^{\rm xc}(\vn{r}).
\ee
Here, $H_{0}$ contains kinetic energy, scalar potential and SOI. The second term on the
right-hand side describes the exchange interaction in collinear magnets.
$\hat{\vn{M}}$ is a normalized vector that points into the magnetization direction.
$\vn{m}=-\mubo\vn{\sigma}$ is the spin magnetic moment operator,
where $\mubo$ is the Bohr magneton 
and $\vn{\sigma}=(\sigma_x,\sigma_y,\sigma_z)^{\rm T}$ is the
vector of Pauli matrices.
The exchange field
$\Bxc^{\rm xc}(\vn{r})=\frac{1}{2\mu_{\rm
    B}}\left[V^{\rm eff}_{\rm minority}(\vn{r})-V^{\rm eff}_{\rm
    majority}(\vn{r}) \right]$
is given by the difference between the effective potentials of minority and
majority electrons.

When the local spin polarization goes down during ultrafast demagnetization
also the exchange splitting is reduced.
We model this by the time-dependent Hamiltonian
\bege
\label{eq_t_dep_ham}
\begin{aligned}
H(\vn{r},t)&=H_{0}(\vn{r})-\vn{m}\cdot\hat{\vn{M}}\Bxc^{\rm xc}(\vn{r})\eta(t)=\\
&=H_{0}(\vn{r})+\mubo\Bxc^{\rm xc}_{\Vert}(\vn{r})\eta(t)
\end{aligned}
\ee
where $\eta(t)$ describes the reduction of the exchange field and
we defined the 
operator $\Bxc^{\rm xc}_{\Vert}(\vn{r})=\vn{\sigma}\cdot\hat{\vn{M}}\Bxc^{\rm xc}(\vn{r})$.
We assume that the exchange field is reduced linearly in time at a 
rate of $\gamma$, i.e., $\rmd\eta/\rmd t=-\gamma<0$.
This induces an electrical current per unit length of
\bege\label{eq_induced_current}
J_{\alpha}=-L_{\alpha}\frac{\gamma}{A},
\ee
where $A$ is the area of the unit cell and we defined the response coefficients
\bege\label{eq_coefficient}
L_{\alpha}\!=\!
e\mubo\!
\lim_{\omega\to 0}\!
\frac{{\rm Im}G_{v^{\phantom{x}}_{\alpha},\Bxc^{\rm xc}_{\Vert}}^{\rm R}
(\hbar\omega)}{\hbar\omega}.
\ee
Here, $\alpha=x,y,z$ denotes the cartesian component, 
$e>0$ is the elementary positive charge, and
\bege
G_{v^{\phantom{x}}_{\alpha},
\Bxc^{\rm xc}_{\Vert}
}^{\rm R}
\!(\hbar\omega)=
-i\int\limits_{0}^{\infty}\rmd t \rme^{i\omega t}
\left\langle
[
v^{\phantom{x}}_{\alpha}(t),
\Bxc^{\rm xc}_{\Vert}(0)
]_{-}
\right\rangle
\ee
is the retarded correlation-function describing the response of the
velocity operator $\vn{v}$ ($v_{\alpha}$ is the $\alpha$-th cartesian component) to 
perturbations by the 
operator $\Bxc^{\rm xc}_{\Vert}$.

Our choice to define the current density as a current per length in 
Eq.~\eqref{eq_induced_current} is convenient for bilayer systems, where the
current distribution is inhomogeneous along the stacking direction. 
The current density $j_{\alpha}$ (current per area)
used to describe periodic bulk systems in three dimensions
is related to $J_{\alpha}$ (current per length) by
\bege
J_{\alpha}=\int \rmd z \,j_{\alpha}(z),
\ee
where $z$ is the coordinate along the stacking direction, which is 
perpendicular to the bilayer
interface.

The coefficient $L_{\alpha}$ defined 
in Eq.~\eqref{eq_coefficient} describes only the electric current
induced by the dynamical exchange splitting and 
consequently $J_{\alpha}$ given by Eq.~\eqref{eq_induced_current}
contains only this current and misses several 
contributions that we 
discussed in the introduction, namely the current from the conversion
of the superdiffusive spin-current into an interfacial charge current
by the inverse spin Hall effect and the current from the inverse
spin-orbit torque and the tilting magnetization due to the inverse
Faraday effect. The theoretical modelling of these two currents
has been described already 
elsewhere~\cite{seifert_THz_emitter,simultaneous_thz_moke_huisman}.

The details of the laser pulse that triggers ultrafast 
demagnetization do not enter
the expression for the current $J_{\alpha}$ given by Eq.~\eqref{eq_induced_current},
only the rate of change of the exchange field matters.
This is a consequence of our model, Eq.~\eqref{eq_t_dep_ham}, which
only phenomenologically describes the reduction of the exchange splitting
during ultrafast demagnetization but does not model the mechanism of
ultrafast demagnetization itself. In order to apply our theory to a 
given experimental setup it is therefore necessary to determine 
the parameter $\gamma$ from measurements of ultrafast demagnetization.
In section~\ref{sec_results} we will provide an example of this procedure.

For the magnetic bilayer systems considered in this work,
Eq.~\eqref{eq_coefficient} can be rewritten as
\bege\label{eq_longitex_final}
L_{\alpha}=-\frac{e\hbar}{\pi\mathcal{N}}
\!\!\sum_{\vn{k}nm}\!
\frac{\mubo\Gamma^2
{\rm Re}
\left[
\langle
\psi^{\phantom{R}}_{\vn{k}n}
|
v_{\alpha}
|
\psi^{\phantom{R}}_{\vn{k}m}
\rangle
\langle
\psi^{\phantom{R}}_{\vn{k}m}
|
\Bxc^{\rm xc}_{\Vert}
|
\psi^{\phantom{R}}_{\vn{k}n}
\rangle\right]
}{
\left[(\mathcal{E}^{\phantom{R}}_{\rm F}-\mathcal{E}^{\phantom{R}}_{\vn{k}n})^2+\Gamma^2\right]
\left[(\mathcal{E}^{\phantom{R}}_{\rm F}-\mathcal{E}^{\phantom{R}}_{\vn{k}m})^2+\Gamma^2\right]
},
\ee
where the Bloch state $|\psi_{\vn{k}n}\rangle$ and the band energy $\mathcal{E}_{\vn{k}n}$ 
satisfy $H|\psi_{\vn{k}n}\rangle= \mathcal{E}_{\vn{k}n} |\psi_{\vn{k}n}\rangle$ with
the Hamiltonian $H$ defined in Eq.~\eqref{eq_ks_hamiltonian} and we
use the constant broadening $\Gamma$ of the electronic bands
in order to describe disorder. $\mathcal{E}^{\phantom{R}}_{\rm F}$ is
the Fermi energy and $\mathcal{N}$ is the number of $\vn{k}$ points
used to sample the Brillouin zone. $L_{\alpha}$ in
Eq.~\eqref{eq_longitex_final} depends on the magnetization direction $\hat{\vn{M}}$
and is odd in $\hat{\vn{M}}$, i.e., $L_{\alpha}(\hat{\vn{M}})=-L_{\alpha}(-\hat{\vn{M}})$.
Generally, Eq.~\eqref{eq_coefficient} can also contain an even
component, which we do not include in Eq.~\eqref{eq_longitex_final}. 
However, due to mirror and rotational symmetries this
even component vanishes in the bilayer systems
considered in this work. 
 
$L_{\alpha}$ describes the coupling between a polar vector and a scalar and
therefore $L_{\alpha}$ is nonzero only in systems with inversion asymmetry.
Furthermore, $L_{\alpha}$ is zero when SOI is not included in the Hamiltonian.

The broadening $\Gamma$ is related to the lifetime $\tau$ of the
electronic states by the expression $\tau=\hbar/(2\Gamma)$. The lifetime
can be controlled in experiments
by modifying the amount of disorder or the temperature. The electrical
conductivity is also sensitive to the lifetime. Measurements of the electrical
conductivity can be used to extract the lifetime and from this the 
broadening $\Gamma$ needed in our theoretical description.
Eq.~\eqref{eq_longitex_final} is valid in the static limit. 
It describes the electric current
induced by a dynamical exchange splitting correctly
when the condition $\gamma\ll\tau^{-1}$
is satisfied.

\section{Ab initio results and discussion}
\label{sec_results}
In the following we use \textit{ab initio} density functional theory
calculations in order to investigate the electric current driven
by the dynamical exchange splitting in Mn/W(001) and Co/Pt(111)
magnetic bilayers.
Both Pt and W provide large SOI at the interfaces, which is ideal
in order to maximize the effect. In a previous work we studied 
in these two systems already
the spin-orbit torques (SOTs), which also require
SOI and inversion asymmetry and are therefore loosely related to
the effect of interest here.
Laser-induced electric currents have been studied experimentally in
Co/Pt already~\cite{simultaneous_thz_moke_huisman}, while such
experiments have not yet been performed in Mn/W(001). However,
monolayers of Mn on W(001) have already been prepared experimentally
and their magnetic structure has been determined from 
spin-polarized scanning tunneling microscopy~\cite{dmi_mnw_ferriani}.

In our calculations, the Mn/W(001) system is
composed of one monolayer of Mn deposited on 9
atomic layers of W(001) and the Co/Pt(111) bilayer consists of three
atomic layers of Co deposited on 10 atomic layers of Pt(111).
Details of the electronic structure calculations
are given in~\cite{ibcsoit}. 
The ground-state magnetic structure of Mn/W(001) is a 
spin-spiral~\cite{dmi_mnw_ferriani}. However, in the
present work we treat Mn/W(001) as a ferromagnet.
We use Eq.~\eqref{eq_longitex_final} in order to calculate the
response coefficient $L_{\alpha}$, where we employ Wannier interpolation
for computational efficiency~\cite{wannier90,WannierPaper}.
Technical details on the Wannier interpolation are given in~\cite{ibcsoit}.  

\begin{figure}
\includegraphics[width=\linewidth,trim=4cm 4.5cm 4.5cm 4.5cm,clip]{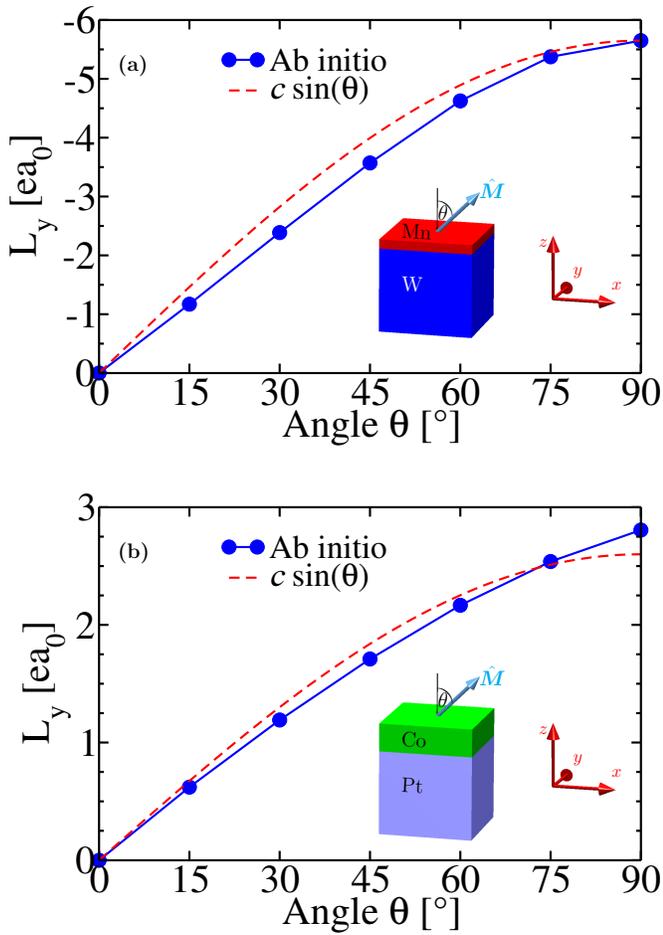}
\caption{\label{figure_longitex} 
Response coefficient $L_y$
vs.\ polar angle $\theta$ of the
magnetization direction in
(a) the Mn/W(001) magnetic bilayer
and
(b) the Co/Pt(111) magnetic bilayer.
Circles: \textit{Ab initio} results.
Dashed line: Fit with $c\sin(\theta)$, where $c$ is the
fitting parameter.
$L_{y}$ is plotted in units of
$ea_{0}=8.478\cdot10^{-30}$Cm, where $e$ is the 
elementary positive charge and $a_{0}$ is Bohr's radius.
The quasiparticle broadening is set to $\Gamma=25$meV.
The insets illustrate the geometry.
}
\end{figure}

In Figure~\ref{figure_longitex} we show the coefficient $L_{y}$ as a function of
the polar angle $\theta$ of the magnetization 
direction $\hat{\vn{M}}=(\sin\theta,0,\cos\theta)^{\rm T}$. The angular 
dependence of $L_{y}$ is approximately given by $\sin\theta$, i.e., $L_{y}$
is roughly proportional to the $x$ component of $\hat{\vn{M}}$. 
The component $L_{x}$ (not shown in the figure) is zero in this case, because
it is proportional to the $y$ component 
of $\hat{\vn{M}}=(\sin\theta,0,\cos\theta)^{\rm T}$, 
which is 
zero. This angular dependence agrees to the one of the inverse 
Edelstein 
effect~\cite{spin_to_charge_rojas_sanchez,inverse_rashba_cubi,PhysRevLett.112.096601},
which is given by $J^{\rm IE}\propto \hat{\vn{z}}\times \dot{\vn{B}}$:
In the case of the inverse Edelstein effect the in-plane electric current  
flowing into $y$ direction is proportional to the rate of change of the $x$ component
of an applied time-dependent magnetic field, or, equivalently, it is proportional to the
$x$ component of the spin injected into the interface due to a spin-current flowing 
towards the bilayer interface. 
The inverse Edelstein effect is a consequence of Rashba-type SOI, which orients
the spin $\vn{S}$ of the electrons 
perpendicularly to 
the $\vn{k}$-vector: $\vn{S}\propto\hat{\vn{z}}\times \vn{k}$.
In the bilayer systems considered in this work the $\vn{k}$-linear term of
the spin-orbit field is given by Rashba SOI. 
Since the charge pumping driven by a dynamical
exchange splitting exhibits the same angular dependence as the
inverse Edelstein effect, it can indeed  be interpreted as a 
ferromagnetic variant of the inverse
Edelstein effect, as anticipated in the introduction. 

\begin{figure}
\flushright
\includegraphics[width=\linewidth,trim=4cm 4.5cm 3cm 5cm,clip]{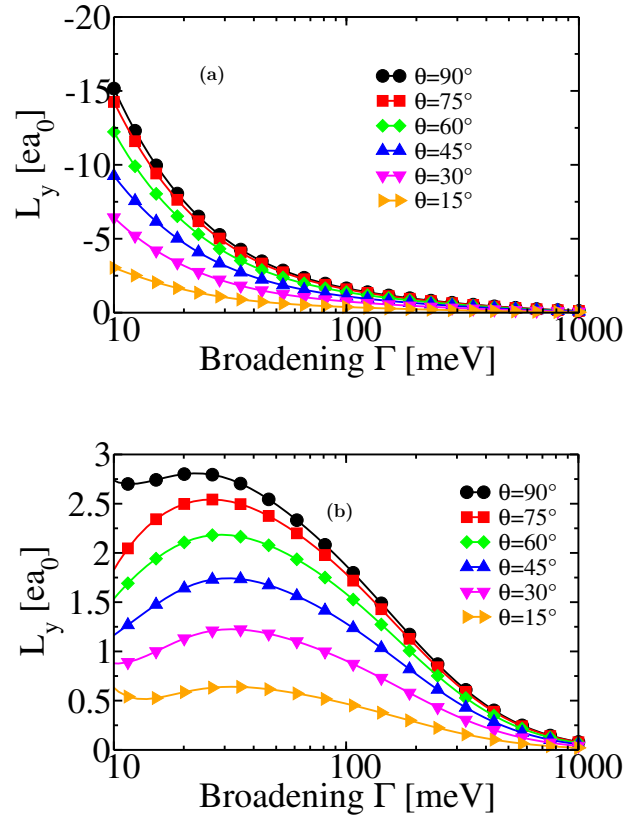}
\caption{\label{figure_longitex_vs_gamma} 
Response coefficient $L_{y}$
vs.\ 
quasiparticle broadening $\Gamma$ in 
(a) the Mn/W(001) magnetic bilayer and
(b) the Co/Pt(111) magnetic bilayer. 
Results are shown for various
polar angles $\theta$ of the magnetization direction.
}
\end{figure}

In Figure~\ref{figure_longitex_vs_gamma} we show $L_{y}$ as 
a function of the quasiparticle broadening $\Gamma$, which we use to model
the disorder. $L_{y}$ depends strongly on $\Gamma$ and vanishes for very large
disorder. Previously, we reported a similarly strong disorder dependence of the
spin-orbit torques (SOTs) in these bilayer systems~\cite{ibcsoit}. 
In the considered $\Gamma$ range from 10~meV to 1000~meV the 
coefficient $L_{y}$ decreases monotonically in Mn/W(001), while it 
exhibits a maximum in Co/Pt(111). A similar qualitative difference in the
$\Gamma$ dependence between Mn/W(001) and Co/Pt(111) is found in the
odd component of the SOT~\cite{ibcsoit}. In order to explain these similarities we
assume that the inverse Edelstein effect 
arises from the interfacial spin-orbit coupling
in these bilayer systems, like the odd component of the SOT.
Additionally, similarly to the odd component of the SOT,
the inverse Edelstein effect is dominated by
intraband terms for small values of $\Gamma$, while interband terms
are activated when $\Gamma$ is increased.

When the magnetization direction lies in the $xz$ 
plane, i.e., when $\hat{\vn{M}}=(\sin\theta,0,\cos\theta)^{\rm T}$,
the $y$ component of the interfacial electric current generated
when a laser-pulse is applied to the system is expected
to consist of two contributions. One contribution arises from the
dynamical exchange splitting and is given 
by $J_{y}=-L_{y}\gamma/A\propto\sin(\theta)\gamma$ 
as discussed above. The second contribution arises from the
conversion of superdiffusive spin currents into interfacial charge currents
by the ISHE~\cite{thz_spin_current_kampfrath}. 
The spin-polarization vector of the superdiffusive spin current
is coupled to the magnetization direction $\hat{\vn{M}}$. Thus, the
amount of spin current converted into charge current $J_{y}$ by ISHE
is proportional to $\sin(\theta)$ like $L_{y}$.
Since the two contributions to $J_{y}$ have therefore the 
same $\theta$ dependence they are difficult to separate experimentally.
In order to reveal a relation between dynamical exchange splitting and charge
currents in experiments unambiguously one possibility is to 
design the experiment
such that the contribution from the superdiffusive spin current is 
suppressed. It seems likely
that this can be achieved in thin magnetic layers grown on insulators.
Alternatively, instead of suppressing the superdiffusive spin current one can
also use magnets with bulk inversion asymmetry and identify the bulk contribution
to the induced electric current. When the half-Heusler NiMnSb is tetragonally strained
the $\vn{k}$-linear term in the spin-orbit field is of Dresselhaus 
symmetry~\cite{Ciccarelli_NiMnSb}. Thereby, one can achieve that the
electric current driven by the dynamical exchange splitting has a different
angular dependence than the electric current from the conversion of the
superdiffusive spin current. For example, one can achieve that the current
driven by the dynamical exchange splitting flows in $x$ direction, 
i.e., $J_{x}=-L_{x}\gamma/A\propto\sin(\theta)\gamma$, 
when $\hat{\vn{M}}=(\sin\theta,0,\cos\theta)^{\rm T}$. 

In a recent experiment the THz emission from
CoFeB/HM magnetic bilayers under laser illumination 
was measured for several different 
choices of the heavy metal (HM) layer, namely Cr, Pd, Ta, W, Ir 
and Pt~\cite{seifert_THz_emitter}.
Thereby, it was found that when $\theta=90^{\circ}$ the interfacial electric current
in $y$ direction, $J_{y}$, is proportional to the spin Hall conductivity
and results from the excitation of superdiffusive spin currents that are
converted into in-plane charge currents by the ISHE.
This interpretation is also consistent with the Pt-thickness 
dependence of $J_{y}$, which suggests that $J_{y}$ is not a pure
interface effect but that the Pt-thickness dependence is determined by the
spin diffusion length, which governs the decay of the 
superdiffusive spin current in the 
HM layer~\cite{seifert_THz_emitter,Huisman_2016}.
Since the electric current driven by the dynamical exchange splitting is not
related to the ISHE, these experimental observations imply
that the contribution from the dynamical exchange splitting to $J_{y}$
must be at least one order of magnitude smaller than the contribution
from the superdiffusive spin current and the ISHE, because otherwise it
would have been observed in these experiments.

Based on our \textit{ab initio} results for the coefficient $L_{y}$ we can
assess the expected order of magnitude of the electric current induced
by a dynamical exchange splitting for various scenarios of demagnetization. 
For 50~fs laser pulses with central wavelength 800~nm and 
fluence 1~mJ/cm$^2$ we 
estimate $\gamma\approx 2\cdot 10^{11}$~s$^{-1}$ in Co from the 
magnetization dynamics extracted from MOKE 
measurements~\cite{simultaneous_thz_moke_huisman_gamma} when we
assume the scenario that no transverse fluctuations are excited
and that the demagnetization corresponds to a reduction of the
local magnetic moments. Previously, we found that a broadening 
of $\Gamma$=25~meV can be used to simulate room-temperature 
SOT experiments on Co/Pt bilayer systems~\cite{ibcsoit}. At
$\Gamma$=25~meV and $\theta=90^{\circ}$ we 
obtain $L_{y}$=2.8~$ea_{0}$
in Co/Pt(111) in our calculations. In this scenario of demagnetization 
the resulting current per length
driven by the dynamical exchange splitting 
would be $J_{y}=-71.35$~A/m, which is even larger than the current 
$J_{y}=32$~A/m
measured
experimentally in Co(10nm)/Pt(2nm) for the laser pulse parameters
given above~\cite{Huisman_2016}.
Since the experimentally observed current $J_{y}$ has been demonstrated to arise 
dominantly from superdiffusive spin currents combined with 
ISHE~\cite{thz_spin_current_kampfrath,seifert_THz_emitter}, we
conclude that the assumption that ultrafast demagnetization in Co 
corresponds to a reduction of the local magnetic moments without
excitation of transverse fluctuations has to be incorrect. This finding
is consistent with the experimental observation that electric conductivity
is not strongly modified during ultrafast demagnetization in Fe(001)/MgO(001),
while any strong reduction of the exchange splitting is expected to result
in a significant variation of electric conductivity~\cite{collapsed_vs_collective}.
Our finding is also consistent with theoretical work on ultrafast
demagnetization, which reports that it is dominated by transverse spin 
fluctuations~\cite{ufd_pastor}.

The electric current driven by transverse spin fluctuations 
has been estimated to be much smaller than the electric current
from the conversion of the superdiffusive 
spin current~\cite{thz_spin_current_kampfrath}.
If we assume the scenario that 
ultrafast demagnetization is dominated by spin fluctuations and that
the reduction of the local magnetic moments contributes  
at most 5\% to the total demagnetization the estimated reduction rate of
the exchange splitting is at most 5\% of $2\cdot 10^{11}$s$^{-1}$, 
i.e., $\gamma<10^{10}$s$^{-1}$ in Co for the laser parameters
given above. In this scenario of demagnetization the electric current $J_{y}$
driven by the dynamical exchange splitting is smaller than 3.6~A/m
in Co/Pt(111) according to our calculations, which is
one order of magnitude smaller than the experimentally measured  $J_{y}$.
Such a small contribution to $J_{y}$ from the
dynamical exchange splitting is consistent with the experimental observation
that $J_{y}$ is clearly dominated by the contribution 
from the superdiffusive spin current converted by the ISHE.

For the laser-pulse parameters used in the experiment, i.e., 
duration 50~fs and fluence 1~mJ/cm$^2$, we can thus
rule out 
by comparing our theoretical calculations to experiment
that ultrafast demagnetization in Co dominantly arises from a reduction of the
modulus of the local magnetic moments without excitation of transverse fluctuations.
For shorter laser pulse duration and higher intensities a reduction of the modulus
of the local magnetic moments without excitation of transverse fluctuations has
been predicted from time-dependent density functional theory 
studies~\cite{demagnetization_tddft}. By performing measurements of the
THz radiation under laser excitation with shorter laser pulses and higher
intensities one should therefore be able to observe a contribution to the THz signal
from a dynamical exchange splitting.

\section{Summary}
\label{sec_summary}
We demonstrate that
a dynamical exchange splitting induces 
measurable electric currents in 
magnetic bilayer systems with inversion asymmetry and
spin-orbit interaction. 
Using an \textit{ab initio} approach, we study this
effect in Mn/W(001) and Co/Pt(111) magnetic bilayers.
The strong disorder-dependence is reminiscent of the odd 
component of the spin-orbit torque in these magnetic bilayers
pointing at the interfacial spin-orbit interaction as common
mechanism. The dependence on magnetization direction
suggests to view electric currents driven by a dynamical
exchange splitting as a ferromagnetic variant of the
inverse Edelstein effect. We compare our theoretical results
to experiments measuring the THz emission from magnetic
bilayer systems under laser illumination. This leads us to
the conclusion that when ultrafast demagnetization in Co
is triggered by 800~nm 50~fs laser pulses this ultrafast
demagnetization is not dominated by a reduction of the
local magnetic moments, suggesting an important role of
transverse spin fluctuations.
  
\section*{Acknowledgments}
We gratefully acknowledge computing time on the supercomputers
of J\"ulich Supercomputing Center and RWTH Aachen University
as well as financial support from the programme
SPP 1538 Spin Caloric Transport
of the Deutsche Forschungsgemeinschaft.

\bibliography{longitex}

\begin{thebibliography}{38}
\bibitem[{sim()}]{simultaneous_thz_moke_huisman_gamma}
\bibinfo{note}{According to Ref.~\cite{simultaneous_thz_moke_huisman} the
  relative change of magnetization in Co amounts to $\Delta M/M\approx 0.1$
  after 0.5ps. Hence we estimate $\gamma\approx 0.1/$(0.5ps)$=2\cdot
  10^{11}$Hz.}
\end{thebibliography}

\end{document}